\documentclass[12pt]{iopart}
\usepackage{iopams}

\usepackage{hyperref}

\expandafter\let\csname equation*\endcsname\relax
\expandafter\let\csname endequation*\endcsname\relax

\usepackage{amsmath}
\usepackage{dsfont}

\usepackage{graphicx}

\usepackage{color}

\begin{document}

\title[Long-time asymptotic series for the Painlev\'e II equation]
{Long-time asymptotic series for the Painlev\'e II equation: Riemann--Hilbert approach}

\author{N Iorgov$^{1,2}$ and Yu Zhuravlov$^{1,2}$}

\address{$^1$ Bogolyubov Institute for Theoretical Physics,  14B Metrolohichna st.,  03143 Kyiv, Ukraine}
\address{$^2$ Kyiv Academic University, 03142 Kyiv, Ukraine}

\eads{\mailto{iorgov@bitp.kyiv.ua}, \mailto{ujpake@gmail.com}}

\vspace{10pt}

\begin{abstract}
We elaborate a systematic way to obtain higher order contributions in the nonlinear steepest descent method for  Riemann--Hilbert problem associated with homogeneous  Painlev\'e~II equation. The problem is reformulated as a matrix factorization problem on two circles
and can be solved perturbatively reducing it to finite systems of algebraic linear equations.
The method is applied to find explicitly long-time asymptotic behaviour for tau function of Painlev\'e~II equation.
\end{abstract}

%
%
%
%
%

\section{Introduction}

Painlev\'e equations are nonlinear differential equations appearing in different problems of physics and mathematics 
\cite{McCoy76,Jimbo1980Density,Douglas1990,Tracy1994Airy,Tracy1994Fredholm}.
These  equation can be obtained as equations governing isomonodromy preserving deformations of linear matrix differential equations on Riemann sphere \cite{Flaschka1980,Jimbo1981}. They can be interpreted as Hamiltonian equations with the Hamiltonians related to so-called  isomonodromic tau functions introduced by Jimbo, Miwa, and Ueno for differential equations with irregular singularities \cite{JMU81}.

From the point of view of asympototic analysis of the solutions of Painlev\'e equations, a reformulation of the problem 
as a Riemann-Hilbert (RH) problem allows one to use the nonlinear steepest descent method   \cite{DeiftZhou93,fokas2006painleve}.
In \cite{Its_2018, Bonelli_2017} a conjecture for large time asymptotic series for tau function of Painlev\'e II equation is conjectured.
The leading terms of this series were obtained in \cite{Kapaev1992,Deift1995,fokas2006painleve} using nonlinear steepest descent method for the corresponding RH problem. 

The main goal of this paper is  to elaborate a systematic way to calculate higher orders of approximation (next to leading terms in asymptotic expansion for $t\to -\infty$) for the Hamiltonian and reobtain the tau function expansion conjectured in \cite{Its_2018, Bonelli_2017}. The idea is to reformulate RH problem as a matrix factorization problem on two circles neglecting exponentially 
small terms. This factorization problem can be solved perturbatively reducing it to finite systems of algebraic linear equations.
The method is applied to find explicitly first few terms of  long-time asymptotic behaviour for tau function of Painlev\'e~II equation.
They coincide with the leading terms of conjectural asymptotic series found  in   \cite{Its_2018, Bonelli_2017}.
Let us mention another paper \cite{Deift1994}, where a different method of finding higher order asymptotics of solutions of MKdV equation 
was proposed. In the recent paper \cite{Deao2023}, RH approach was adopted to find exponentially small correction 
for long-time asymptotic expansion of tronqu\'ee solutions of Painlev\'e~I equation.

The paper is organized as follows. In section 2 we recall the definition of 
homogeneous  Painlev\'e~II equation, its relation to isomonodromy preserving transformation of an associated linear problem, 
Jimbo--Miwa--Ueno isomonodromic tau function and its asymptotics for long time. 
In section~3 we formulate a Riemann--Hilbert problem 
associated with the Painlev\'e~II equation preparing the ground for the application of nonlinear steepest descent method.
This section follows \cite{fokas2006painleve} closely.
Section~4 describes how to reduce the Riemann--Hilbert problem to the matrix factorization problems on two circles neglecting
exponentially small terms. Also there we give a definition of degree of approximation used in the paper.
In sections 5 and 6 we show how to build solutions of the factorization problems perturbatively reducing such problems to solving finite systems of algebraic linear equations. To ease the reading we explain the procedure by explicit calculations  for the first order
of approximation. A short summary and outlook are presented in section 7.
Appendices contain some technical details concerning factorization problems.

\section{Painlev\'e~II equation and related linear problem}

\subsection{Painlev\'e~II equation  and its Hamiltonian formulation}
In this paper we will study the homogeneous Painlev\'e~II equation defined by
\begin{equation}\label{PIIeq}
	\frac{d^2 q}{dt^2}=2q^3+tq.
\end{equation}
It is the equation of motion of a Hamiltonain system with the time dependent Hamiltonian
\begin{equation}
    H(p,q,t)=\frac{p^2}{4}-q^4-tq^2.
\end{equation}
In what follows we will use expression for the Hamiltonian on trajectories
\begin{equation}
    H(t)=q_t^2-q^4-tq^2.
\end{equation}
The central object of our study is the isomonodromic tau function $\tau(t)$ defined by
\begin{equation}\label{taudef1}
    \partial_t \log \tau(t)=H(t).
\end{equation}

\subsection{Isomonodromic reformulation of Painlev\'e~II equation}
There is another way to obtain the Painlev\'e~II equation. Consider the following system of differential equations, called the linear problem
\begin{equation}\label{linprob}
	\partial_z \Phi=A(z)\Phi, \qquad \partial_t \Phi=B(z)\Phi,
\end{equation}
with $2\times 2$ traceless matrices $A(z)$ and $B(z)$ defined by
\begin{equation} 
    A(z)=\begin{pmatrix}
        -4i z^2 -it - 2iq^2 &  4izq-2q_t\\
      -4izq-2q_t   & 4i z^2 +it +2iq^2
    \end{pmatrix}, \quad B(z)=\begin{pmatrix}
        -iz & iq\\
        -iq & iz
        \end{pmatrix}.
\end{equation}
Imposing zero curvature condition we get the Painlev\'e II equation
\begin{equation}
	\partial_t A -\partial_z B +[A,B]=-2\left(\frac{d^2 q}{dt^2}-2q^3-tq\right)\sigma_1=0,
\end{equation}
where we use standard Pauli matrices
\begin{equation}
		\sigma_1=\begin{pmatrix}
			0 & 1\\
			1 & 0
		\end{pmatrix}, \quad
		\sigma_2=\begin{pmatrix}
			0 & -i\\
			i & 0
		\end{pmatrix}, \quad
		\sigma_3=\begin{pmatrix}
		1 & 0\\
		0 & -1
		\end{pmatrix}.
\end{equation}
The linear problem \eref{linprob} has irregular singular point at $z=\infty$ and can be solved formally for large $z$ using the following anzats
\begin{equation}\label{formalas}
    \Phi(z)=\Psi(z)e^{\Theta(z)}, \quad \Theta(z)=-i\left(\frac{4z^3}{3}+tz\right)\sigma_3,
\end{equation}
with  formal matrix series (here $I$ is $2\times 2$ identity matrix)
\begin{equation}\label{psias}
    \Psi(z)=I+\sum_{k=1}^{\infty}g_k z^{-k}.
\end{equation}
The exponent $\Theta(z)$ follows from the large $z$ contribution of the integral 
\begin{equation}
    \int^z dz \sqrt{\frac{1}{2}\Tr A(z)^2} \,\sigma_3= \Theta(z)+O\left(z^{-1}\right),
\end{equation}
where
\begin{equation}
    \frac{1}{2}\Tr A(z)^2=-16z^4-8t z^2 -t^2 + 4H(t).
\end{equation}
The coefficients $g_k$ of the expansion \eref{psias} can be found  from the linear problem \eref{linprob} recursively. We present here first two of them
\begin{equation}\label{asympexpansion}
    g_1=\frac{q}{2}\sigma_1-\frac{iH}{2}\sigma_3, \quad g_2=\frac{1}{8}(q^2-H^2)I-\frac{1}{4}(q_t+qH)\sigma_2.
\end{equation}
The matrix structure of these coefficients can be guessed from the symmetry of the problem
\begin{equation}\label{symrel}
    A(-z)=-\sigma_2 A(z) \sigma_2, \quad \Theta(-z)=\sigma_2 \Theta(z)\sigma_2
\end{equation}
without direct computations. 
These two symmetry relations imply that the formal  matrix series $\Psi(z)$ given by \eref{psias} should satisfy  $Z_2$-symmetry termwise
\begin{equation}\label{psisymrel}
    \Psi(-z)=\sigma_2 \Psi(z) \sigma_2. 
\end{equation}

\subsection{Jimbo--Miwa--Ueno tau function and its large time ($t\to-\infty$) asymptotics}

The formal asymptotic series \eref{formalas} can be used to define Jimbo--Miwa--Ueno (JMU) tau function by the following integral
\begin{equation}\label{tauJMU}
    \partial_t \log \tau_{\mathrm{JMU}}(t)=\oint_{C}\frac{dz}{2\pi i}\mathrm{Tr}\,\Psi(z)^{-1}\partial_z \Psi(z)\partial_t \Theta(z)=H(t),
\end{equation}
where the contour $C$ is a counterclockwise oriented circle of large radius. We see that the tau function defined in \eref{taudef1} coincides with JMU tau function. 
There is a conjecture \cite{Its_2018, Bonelli_2017} about asymptotic behaviour of tau function \eref{tauJMU} for large negative $t$ 
\begin{equation}\label{tausum}
    \tau(t)=\sum_{n=-\infty}^{\infty}e^{in\rho}Z(\nu+n,t),
\end{equation}
where the function $Z(\nu,t)$ is defined by
\begin{equation}\label{sdef}
    Z(\nu,t)=C(\nu) e^{i\nu s}s^{-\nu^2}\left(1+\sum_{k=1}^{\infty}\frac{B_k(\nu)}{s^k}\right), \quad s=\frac{4}{3}(-t)^{\frac{3}{2}},
\end{equation}
with the coefficient $C(\nu)$ independent of $t$  
\begin{equation}
    C(\nu)=6^{-\nu^2}e^{\frac{i\pi\nu^2}{2}}G(1+\nu)^2,
\end{equation}
where $G(x)$ is the Barnes $G$-function.
The coefficients $B_k(\nu)$ are polynomials in $\nu$
\begin{equation}
    B_1(\nu)=-\frac{i\nu(34\nu^2+1)}{18}, \quad B_2(\nu)=-\frac{\nu^2(1156\nu^4+2318\nu^2+271)}{648}, \quad \ldots
\end{equation}
Two parameters $\nu$ and $\rho$ in this formula are related to the monodromy data of the problem \eref{linprob} which will be described in the next section, see \eref{nudef} and \eref{rhodef}.  

\section{Reformulation as a Riemann--Hilbert problem}

\subsection{Stokes graph, canonical solutions in Stokes sectors, Stokes matrices}
\begin{figure}[h!]
	\centering
	\includegraphics{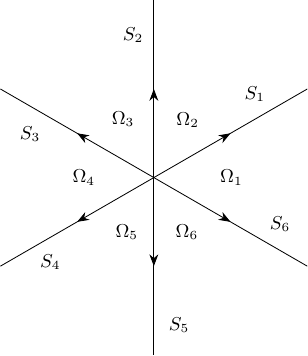}
 \caption{Stokes graph for the original RH problem  (centered at $z=0$)}
 \label{SGPII}
\end{figure}

The linear problem \eref{linprob} can be reformulated as a Riemann--Hilbert (RH) problem \cite{fokas2006painleve}.
Solutions $\Phi(z)$ of \eref{linprob} 
exhibit Stokes phenomenon near irregular singular point $z=\infty$. To describe the phenomenon we plot Stokes graph which follows from the leading term of $\Theta(z)$, namely, the  Stokes rays (see Figure~\ref{SGPII}) are fixed by the relation
\begin{equation}
	\mathrm{Im} \left(-iz^3\right)=0 \qquad \Rightarrow \qquad \mathrm{Arg}\, z=-\frac{\pi}{6}+\frac{\pi k}{3}, \quad k\in \mathbb{Z}.
\end{equation}
It is known that in each Stokes sector $\Omega_k$, enclosed by Stokes rays, there is unique solution, so called canonical solution, $\Phi_k(z)$ 
of  \eref{linprob} having asymptotic behavior  \eref{formalas} as $z\to \infty$ and $z\in \Omega_k$.
Any two solutions of the linear problem \eref{linprob} are related by right multiplication on a constant non-degenerate matrix. 
In particular, the canonical solution $\Phi_k(z)$,  analytically continued to the neighbor Stokes sector $\Omega_{k+1}$, is related to  
$\Phi_{k+1}(z)$ by a Stokes matrix $S_k$: 
 \begin{equation}
	\Phi_{k+1}(z)=\Phi_k(z)S_k,
\end{equation}
where $S_k$ are Stokes matrices having the following form
\begin{equation}
S_{2j+1}=\begin{pmatrix}
		1 & 0\\
		s_{2j+1} & 1
	\end{pmatrix}, \qquad 	
 S_{2j}=\begin{pmatrix}
	1 & s_{2j}\\
	0 & 1
	\end{pmatrix}. 
\end{equation}
The symmetry relation \eref{symrel} implies that 
\begin{equation}\label{symrelStokes}
	S_{k+3}=\sigma_2 S_k \sigma_2 \qquad \Rightarrow \qquad s_{k+3}=-s_k.
\end{equation}
Since $A(z)$ has no singularity at $z=0$, we obtain that analytically continued solution $\Phi(z)$ of \eref{linprob} has no monodromy around $z=0$ (and $z=\infty$): 
$\Phi(e^{2\pi i}z)=\Phi(z)$. Adding the fact that the exponent $\Theta(z)$ does not have a formal monodromy (i.e. $\Theta(z)$ is a single-valued function near $z=\infty$), 
we obtain the cyclic relation on Stokes matrices
\begin{equation}\label{cycrel}
	S_1 S_2 S_3 S_4 S_5 S_6 = 1 \qquad \Rightarrow \qquad s_1-s_2+s_3+s_1 s_2 s_3=0.
\end{equation}
This shows that there are only two independent Stokes parameters. It is convenient to define  parameter $\nu$ by 
\begin{equation}\label{nudef}
    e^{2\pi i\nu}=1-s_1 s_3, \qquad |\mathrm{Re}\,\nu|<\frac{1}{2},
\end{equation}
and parameter $\rho$
\begin{equation}\label{rhodef}
    e^{i\rho}=\frac{s_3^2}{2\pi(1-s_1 s_3)}.
\end{equation}
The monodromy parameters $\nu$ and $\rho$ enter the asymptotic expansion of tau function \eref{tausum}.
Instead of $\rho$ we will also use $Q$ defined by
\begin{equation}\label{Qdef}
    e^{i\rho}=\frac{Q^2e^{-4i\pi \nu}}{\Gamma^2(\nu)}.
\end{equation}

In order to formulate  a RH problem associated with \eref{linprob} we use matrix functions
\begin{equation}\label{Psik1}
    \Psi_k(z)=\Phi_k(z)e^{-\Theta(z)}, \qquad z\in \bar \Omega_k, \quad k=1,2,\ldots,6,
\end{equation}
which are related by the jump matrices $G_k(z)$ on the Stokes rays:
\begin{equation}\label{jumpdefRH1}
	\Psi_{k+1}(z)=\Psi_k(z) G_k(z), \qquad G_k(z)=e^{\Theta(z)}S_k e^{-\Theta(z)}, \qquad z\in \ell_k.
\end{equation}
Let us introduce {\em piecewise} holomorphic matrix function $\Psi(z)$  defined by 
\begin{equation}
    \Psi(z)= \Psi_k(z), \qquad z\in \Omega_k, \quad k=1,2,\ldots,6.
\end{equation}
Then $\Psi(z)$ solves RH problem on the Riemann sphere with RH graph coinciding with Stokes graph and with jump on the graph given by
\begin{equation}\label{RHjump}
	\Psi_+(z)=\Psi_-(z) G_k(z), \qquad z\in \ell_k,
\end{equation}
where $\Psi_+(z)$ (resp. $\Psi_-(z)$) are limiting boundary value of $\Psi(z)$ on RH graph from left (resp. right) with respect to the direction of 
graph edge $\ell_k$.
Finally, the associated RH problem is given by the jump condition  \eref{RHjump}, which, together with the normalization condition 
\begin{equation}
\Psi(z) = I+ O(z^{-1}), \qquad z\to \infty,
    \label{RHnorm}
\end{equation}
fixes the solution uniquely.

Note, $G_k(z)\to I$ exponentially fast as  $z\to \infty$ along $z\in \ell_k$. This observation will play an important role  in the nonlinear steepest descent method 
for large time asymptotic analysis. In the next subsection we reformulate RH problem by deforming RH graph in order to apply the method.

\subsection{Transformation of Stokes graph for nonlinear steepest descent method}

The main idea of the nonlinear steepest descent method \cite{DeiftZhou93} is to solve RH problem near critical points and then to glue all local solutions together. Thus we need to find these critical points. To do this we make rescaling of variables
\begin{equation}\label{rescalingvar}
	z=(-t)^{1/2}\lambda, \qquad x=(-t)^{3/2}=\frac34 s,
\end{equation}
then the exponent $\Theta$ becomes
\begin{equation}\label{thetathetarel}
	\Theta(z,t)=-i\left(\frac{4z^3}{3}+tz\right)\sigma_3=x\theta(\lambda)\sigma_3, \qquad \theta(\lambda)=-i\left(\frac{4\lambda^3}{3}-\lambda\right).
\end{equation}
The critical points are defined by $\theta'(\lambda)=0$ giving $\lambda_\pm=\pm 1/2$.
\begin{figure}[h!]
	\centering
	\includegraphics{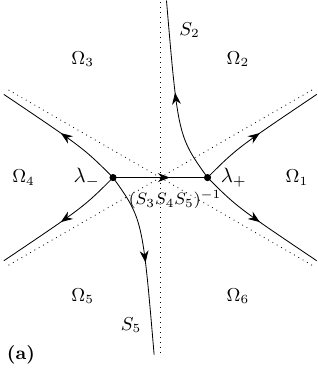} 
 \quad
 \includegraphics{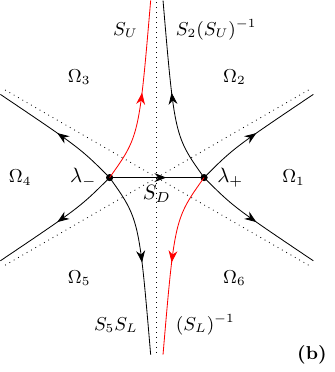}
 \caption{Deformation of the original RH problem.}
 \label{deformedRH}
\end{figure}

On the next step we deform the Stokes graph, see Figure~\ref{deformedRH}. The first deformation allows to go from $\Omega_3$ to $\Omega_6$ with jump on the bridge $(S_3 S_4 S_5)^{-1}$. The second deformation based on the observation that jump matrix on the bridge can be LDU-decomposed
\begin{equation}
	(S_3 S_4 S_5)^{-1}=\begin{pmatrix}
		1-s_1 s_3 & s_1\\
		s_1 & 1+s_1 s_2
	\end{pmatrix}=S_L S_D S_U,
\end{equation}
where we used
\begin{equation}\label{LDUdecomp}
    S_L=\begin{pmatrix}
	1 & 0\\
	\frac{s_1}{1-s_1 s_3} & 1
	\end{pmatrix}, \quad
    S_D=\begin{pmatrix}
	1-s_1 s_3 & 0\\
	0 & (1-s_1 s_3)^{-1}
	\end{pmatrix}, \quad
    S_U=\begin{pmatrix}
	1 & \frac{s_1}{1-s_1 s_3}\\
	0 & 1
	\end{pmatrix}.
\end{equation}
New Stokes matrices $S_U$, $S_2 (S_U)^{-1}$, $S_L^{-1}$, and $S_5 S_L$ are related by $Z_2$-symmetry \eref{symrelStokes} 
\begin{equation}
	S_U=\sigma_2 S_L^{-1}\sigma_2=\begin{pmatrix}
		1 & \frac{s_1}{1-s_1 s_3}\\
		0 & 1
	\end{pmatrix}, \quad S_2 (S_U)^{-1}=\sigma_2 S_5 S_L\sigma_2=\begin{pmatrix}
	1 & \frac{s_3}{1-s_1 s_3}\\
	0 & 1
	\end{pmatrix}.
\end{equation}

\subsection{External local parametrix}
Before we find local solutions near the critical points we build an external parametrix, which is defined as a solution to the RH problem with only one jump $S_D$ on the interval $(\lambda_-,\lambda_+)$, which we will call {\em bridge}. It will help us to glue local solutions. 
The jump matrix $S_D$ on the bridge can be written as
\begin{equation}
    S_D=e^{2\pi i\nu\sigma_3},
\end{equation}
where $\nu$ is defined in \eref{nudef}.
A solution of this auxiliary RH problem is
\begin{equation}\label{psiDdef}
    Y_D(\lambda)=\left(\frac{\lambda-\lambda_+}{\lambda-\lambda_-}\right)^{\nu\sigma_3}.
\end{equation}
It is normalized, i.e. $Y_D(\lambda)\to I$ as $\lambda\to\infty$.

\subsection{Local parametrices near saddle points}

In this subsection, we will build local solutions of the RH problem in disks $D_r$ and $D_l$ of fixed radius $0<r_0<1/2$ centered at critical points $\lambda_+$ and $\lambda_-$, respectively. 
We start from the building conformal maps of these disks to the complex plane straightening the Stokes graph restricted to these disks.
They will be given in terms of the functions $w_\pm(\lambda)$ defined by 
\begin{equation}\label{wpdef}
	w_{\pm}^2(\lambda)=\frac{\theta(\lambda)-\theta(\lambda_{\pm})}{\frac{1}{2}\theta''(\lambda_\pm)}.
\end{equation}
The functions  $w_\pm(\lambda)$ near $\lambda_\pm$, respectively, have the following leading expansion terms
\begin{equation}
	w_\pm(\lambda)=(\lambda-\lambda_\pm)+O((\lambda-\lambda_\pm)^2).
\end{equation}
The conformal map of the right disk $D_r$ is given by
\begin{equation}\label{zetadef}
	\zeta(\lambda)=c_+ x^{\frac{1}{2}}w_+(\lambda), \qquad c_+=\sqrt{2\theta''(\lambda_+)}=\sqrt{8}e^{3\pi i/4},
\end{equation}
where the multiplier $c_+$ is chosen to have
\begin{equation}\label{zetathetarel}
	\frac{\zeta^2(\lambda)}{4}=x\left(\theta(\lambda)-\theta(\lambda_+)\right).
\end{equation}
Similarly, the conformal map of the left disk $D_l$ is given by $\xi(\lambda)=\zeta(-\lambda)$:
\begin{equation}
	\xi(\lambda)=c_- x^{1/2}w_-(\lambda), \qquad c_-=\sqrt{-2\theta''(\lambda_-)}=\sqrt{8}e^{-i\pi/4}.
\end{equation}
The function $\xi(\lambda)$ satisfies
\begin{equation}
	\frac{\xi^2(\lambda)}{4}=-x(\theta(\lambda)-\theta(\lambda_-)).
\end{equation}

To build local solution of the RH problem in $D_r$ and $D_l$,  we consider a model RH problem associated with a linear problem for Weber--Hermite functions
\begin{equation}\label{branchedWH}
	\partial_\zeta Y_{\rm WH}(\zeta)=A_{\rm WH}(\zeta)Y_{\rm WH}(\zeta), \qquad A_{\rm WH}(\zeta)=\begin{pmatrix}\zeta/2 & \nu \\
		1 & -\zeta/2
		\end{pmatrix}.
\end{equation}
Asymptotic behavior of its solutions near the irregular singular point $\zeta=\infty$ is given by
\begin{equation}\label{ydef}
	Y_{\rm WH}(\zeta) \simeq G_0(\zeta)e^{\Theta_{\mathrm{WH}}(\zeta)}, \qquad \Theta_{\mathrm{WH}}(\zeta)=\left(\frac{\zeta^2}{4}+\nu\log \zeta\right)\sigma_3,
\end{equation}
up to a constant matrix multiplier from the right.
The formal series $G_0(\zeta)$ has the form
\begin{equation}\label{G0def}
	G_0(\zeta)=I+\sum_{j=1}^\infty \frac{\rho_j}{\zeta^j}
\end{equation}
 with the  matrices $\rho_j$ defined by
\begin{equation}
	\rho_{2k+1}=\frac{1}{2^k k!}\begin{pmatrix}
		0 & (-1)^{k-1} (\nu)_{2k+1}\\
		(1-\nu)_{2k}& 0
	\end{pmatrix}, 
\end{equation}
\begin{equation}
\rho_{2k}=\frac{1}{2^k k!} \begin{pmatrix}
            (-\nu)_{2k} & 0\\
		0 & (-1)^k (\nu)_{2k}\end{pmatrix},
\end{equation}
where $(x)_j$ is the Pochhammer symbol. 

\begin{figure}[h!]
	\centering
	\includegraphics{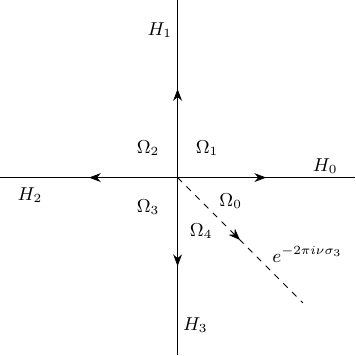}
 \caption{Model RH problem for the local solution near $\lambda_+$.}
 \label{figRHWH}
\end{figure}

Using the exponent $\Theta_{\mathrm{WH}}$ we can build the Stokes graph of the linear problem, see Figure~\ref{figRHWH}. It has four Stokes rays. 
In each sector $\Omega_j$, $j=0,1,\ldots,4$, there is unique solution $Y_j(\zeta)$ of \eref{branchedWH}, so called {\em the canonical solution} in the sector $\Omega_j$,  
which has the asymptotic expansion of form \eref{ydef}. 
We combine all these canonical solutions to {\em piecewise} holomorphic matrix function $Y(\zeta)$ 
\begin{equation}
    Y(\zeta)= Y_j(\zeta), \qquad \zeta \in \Omega_j,\quad j=0,1,\ldots,4,
\end{equation}
having constant jumps on each of Stokes rays:
\begin{equation}
    Y_{j+1}(\zeta)=Y_{j}(\zeta)H_j.
\end{equation}
From the symmetry of the problem \eref{branchedWH} and regularity of $A(\zeta)$ at $\zeta=0$ we can deduce the form of the Stokes matrices
\begin{equation}
	H_{2k}=\begin{pmatrix}
	    1 & 0 \\ h_{2k} & 1
	\end{pmatrix},\quad
	H_{2k+1}=\begin{pmatrix}
	    1 & h_{2k+1} \\ 0 & 1
	\end{pmatrix}.
\end{equation}
They satisfy the relations
\begin{equation}
    H_{k+2}=\sigma_3 e^{-i\pi\nu\sigma_3}H_k e^{i\pi\nu\sigma_3}\sigma_3, \quad H_0 H_1 H_2 H_3=e^{2\pi i\nu\sigma_3}
\end{equation}
implying the relation for the Stokes parameters 
\begin{equation}\label{h0h1nu}
    1+h_0 h_1 = e^{-2\pi i\nu}.
\end{equation}
Explicit formulas for $h_0$ and $h_1$ can be derived from the canonical solutions $Y_j(\zeta)$ written in terms of Weber--Hermite functions:
\begin{equation}
	h_0=-\frac{i\sqrt{2\pi}}{\Gamma(1-\nu)},\qquad
	h_1=\frac{\sqrt{2\pi}}{\Gamma( \nu)}\,e^{-i\pi\nu}.
\end{equation}

In order to construct a local parametrix near $\lambda_+$ and fit Stokes matrices, we 
introduce new {\em piecewise} holomorphic function $\tilde{Y}(\zeta)$ by 
\begin{equation}\label{ytdef}
   \tilde{Y}(\zeta)=Y(\zeta)Q^{-\sigma_3/2},
\end{equation}
where $Q$ is defined by two equivalent (due to \eref{nudef} and \eref{h0h1nu}) relations (see also \eref{Qdef})
\begin{equation}
	Q^{-1}=\frac{-h_1}{s_3}=-\frac{s_1 e^{-2\pi i\nu}}{h_0}.
\end{equation}
Then the jump matrices $\tilde{H}_j$ of  $\tilde{Y}(\zeta)$ defined by $\tilde{Y}_{j+1}(\zeta)=\tilde{Y}_{j}(\zeta)\tilde{H}_j$, or explicitly
$\tilde H_j=Q^{\sigma_3/2} H_j Q^{-\sigma_3/2}$, are exactly jump matrices on Stokes curves attached to the critical points $\lambda_+$, see Figure~\ref{deformedRH}(b):
\begin{equation}
	\tilde{H}_0=(S_L)^{-1}, \quad \tilde{H}_1=S_6,
\end{equation}
\begin{equation}
	\tilde{H}_2=S_1, \quad \tilde{H}_3=S_2 (S_U)^{-1}, \quad \tilde{H}_D=(S_D)^{-1}.
\end{equation}

The constructed function $\tilde{Y}(\zeta)$ solves RH problem in $D_r$ with constant jump matrices. In order to describe local solutions in $D_r$ with 
``dressed'' jump matrices $G_k$ given by \eref{jumpdefRH1}, it is convenient to use the local parametrix
\begin{equation}\label{psir}
    \psi_r(\lambda) = \tilde{Q}^{\sigma_3/2} Y(\zeta(\lambda))Q^{-\sigma_3/2}e^{-\Theta(\lambda)}, \qquad \lambda \in D_r,
\end{equation}
where
\begin{equation}\label{Qtilde}
    \tilde{Q}=Q e^{\frac{2ix}{3}}\left(c_+ x^{\frac12}\right)^{-2\nu}.
\end{equation}
We will use slightly different form of  $\psi_r(\lambda)$ rewritten as
\begin{equation}\label{psirdef}
    \psi_r(\lambda)=G_r(\lambda)\beta_r(\lambda)^{-1}Y_D(\lambda), \quad \psi_l(\lambda)=\sigma_2 \psi_r(-\lambda)\sigma_2,
\end{equation}
where we used the definition \eref{zetadef} of $\zeta(\lambda)$ and introduced the functions
\begin{equation}\label{Grdef}
    \beta_r(\lambda)=\left(\frac{\lambda-\lambda_+}{\lambda-\lambda_-}\cdot\frac{1}{w_+(\lambda)}\right)^{\nu\sigma_3}, \quad G_r(\lambda)= \tilde{Q}^{\sigma_3/2}G_0(\zeta(\lambda))\tilde{Q}^{-\sigma_3/2}.
\end{equation}
Also, we will use the following form of series expansion for $G_r$
\begin{equation}\label{epsilondef}
    G_r(\lambda)=I+\sum_{j=1}^{\infty}\frac{\varepsilon_j}{w_+^j(\lambda)}, \qquad \varepsilon_j=\tilde{Q}^{\sigma_3/2}\rho_j \tilde{Q}^{-\sigma_3/2} \left(c_+ x^{1/2}\right)^{-j}.
\end{equation}

\subsection{Global RH solution and gauge transformations of local parametrices}
We have constructed local parametrices \eref{psiDdef} and \eref{psir} together with conformal map \eref{zetadef} 
solving RH problem locally on $D_0:=\mathbb{CP}^1\setminus D_r \cup D_l $ and $D_r$, respectively. However, there is a freedom, called gauge transformation, to multiply them from the left by functions from $\mathcal{H}(U)$ --- {\em the space of holomorphic invertible matrix functions on  $U$}, where $U$ in these cases are domains where these local parametrices are defined. Thus, in general, the local parametrices have the  form
\begin{equation}\label{localparamD}
    \Psi_D(\lambda)=B_D(\lambda)Y_D(\lambda), \quad \lambda \in D_0,
\end{equation}
\begin{equation}\label{localparam}
    \Psi_r(\lambda)=B_r(\lambda)\psi_r(\lambda), \quad \lambda \in D_r,
    \qquad 
    \Psi_l(\lambda)=\sigma_2 \Psi_r(-\lambda)\sigma_2, \quad \lambda \in D_l. 
\end{equation}
The last relation follows from $Z_2$-symmetry \eref{psisymrel}, which we want to preserve. The precise form of gauge transformation 
$B_r(\lambda)\in \mathcal{H}(D_r)$ and $B_D(\lambda)\in \mathcal{H}(D_0)$ 
will follow from the factorization procedure for jumps between $\Psi_{r}$, $\Psi_{l}$, and $\Psi_D$ described in the next section. 
Note, that normalization condition \eref{RHnorm} requires $B_D(\infty)=I$.

\section{Approximated RH problem on two circles}
\subsection{General idea}
\begin{figure}[h!]
	\centering
\includegraphics{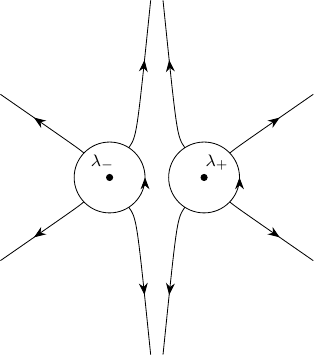} 
 \caption{The RH problem for function $\varphi(z)$.}
\label{twocircleRH}
\end{figure}

 In this section, we will show how to approximate the solution of initial RH problem by the solution of RH problem on two circles.
  We start from the definition of approximated solution $\hat{\Psi}(\lambda)$ for RH problem using parametrices \eref{localparamD} and \eref{localparam} 
\begin{equation}
    \hat{\Psi}(\lambda)=\begin{cases}
    \Psi_D(\lambda), & \lambda\in D_0,\\
    \Psi_r(\lambda), & \lambda\in D_r,\\
    \Psi_l(\lambda), & \lambda\in D_l.
    \end{cases}
\end{equation}
Note that parametrices $\Psi_{r}$ and $\Psi_{l}$ can be analytically continued to larger regions but for now it is sufficient to define them inside disks. We formulate new RH problem for $\varphi(z)$ defined by
\begin{equation}\label{varphidef}
    \varphi(z)=\Psi(z)\hat{\Psi}(\lambda)^{-1}, \qquad \lambda=z/(-t)^{1/2}.
\end{equation}
This new function has jumps on boundary circles $\mathcal{C}_r$ and $\mathcal{C}_l$ of $D_r$ and $D_l$, respectively,  and on tails of Stokes curves, see Figure~\ref{twocircleRH}.
Let us assume that we found parametrices \eref{localparamD} and \eref{localparam}  satisfying  
\begin{equation}
    \Psi_D(\lambda)\Psi_{r}(\lambda)^{-1}=I, \qquad
    \Psi_D(\lambda)\Psi_{l}(\lambda)^{-1}=I
\end{equation}
on circles $\mathcal{C}_r$ and $\mathcal{C}_l$, respectively.
These relations define factorization problems, which allow us to find appropriate gauge transformations $B_D$, $B_r$, and $B_l$. 
For example, for the right circle $\mathcal{C}_r$, we have
\begin{equation}\label{factorprobdef}
    J_r(\lambda):=Y_D(\lambda)\psi_r(\lambda)^{-1}=B_D(\lambda)^{-1}B_r(\lambda),
\end{equation}
where $J_r(\lambda)$ can be rewritten using \eref{psirdef} as
\begin{equation}\label{Jrdef}
    J_r(\lambda)=\beta_r(\lambda)G_r(\lambda)^{-1}.    
\end{equation}
Thus, the  function $\varphi(z)$ defined in \eref{varphidef} has jumps only on tails of the Stokes curves
\begin{equation}\label{expapprox}
    J^{(\mathrm{tail})}_k=\varphi_-(z)^{-1}\varphi_+(z)=\Psi_D(\lambda) e^{\Theta(z)}S_k e^{-\Theta(z)}\Psi_D(\lambda)^{-1}=I+O(e^{-2x|\theta(r_0)|}).
\end{equation}
In this way, the solution $\Psi(z)$ for the initial RH problem can be approximated for large parameter $x$  by $\hat{\Psi}(\lambda)$ up to exponentially small terms. 

In what follows, it is convenient to define another formula for the Hamiltonian
\begin{equation}\label{HHtau}
    \hat{H}=\partial_s \log \tau =\frac{d t}{d s} \cdot \partial_t \log \tau= -\frac{H}{2(-t)^{1/2}},
\end{equation}
where we used the  definition of parameter $s$ from \eref{sdef}. Using the fact that $\Psi(z)$ and $\hat{\Psi}(z)$ are equal up to exponentially small terms, we can relate $\hat{H}$ with the first coefficient of expansion of $\hat{\Psi}$ for large $\lambda$, see also \eref{asympexpansion},
\begin{equation}\label{Hamhatdef}
    \hat{H}\approx -\frac{i}{2}\Tr \hat{g}_1 \sigma_3, \qquad  \hat{\Psi}(\lambda)=I+\sum_{k=1}^{\infty}\frac{\hat{g}_k}{\lambda^k}.
\end{equation}

The strategy of finding large $t$ expansion of tau function using RH approach is to find   $B_D(\lambda)$ by factorization of left-hand side of \eref{factorprobdef} in some order of approximation
(see the precise definition below)
and then  using \eref{localparamD} to obtain an approximation for the global solution $\Psi(\lambda)\approx \hat \Psi(\lambda)$ of RH problem, 
which, in turn, can be used to recover the Hamiltonian $\hat H$ and the tau function $\tau(t)$ in the same order of approximation by means of relations \eref{Hamhatdef} and \eref{HHtau}.

\subsection{Gluing of local solutions of RH problem}
In order to glue local  solutions of RH problem we need to solve the factorization problem \eref{factorprobdef} and its left circle counterpart.
This problem will be solved in two steps. The first step is factorization of jumps on both circles. If such factorizations exist, on the second step we use a freedom of choosing these factorizations to fix the external parametrix.  
Let a solution of factorization problem on the right circle is given by 
\begin{equation}\label{Jrfactor}
    J_r(\lambda)=\left(J_{r,-}(\lambda)\right)^{-1}J_{r,+}(\lambda),
\end{equation}
where $J_{r,-}(\lambda)\in \mathcal{H}(D_0)$ and  $J_{r,+}(\lambda)\in \mathcal{H}(D_r)$, respectively. 
This factorization is not unique. 
Namely, both functions $J_{r,\pm}$ can be multiplied by a function $R_r(\lambda) \in \mathcal{H}(\mathbb{CP}^1\setminus D_l)$ from the left without spoiling the relation. 
For the left circle the story is similar and we have a freedom in choosing factorization describing by $R_l(\lambda)\in \mathcal{H}(\mathbb{CP}^1\setminus D_r)$. Thus  the gauge transformations ${B}_r(\lambda)$, ${B}_l(\lambda)$  are
\begin{equation}
    {B}_r(\lambda)=R_r(\lambda)J_{r,+}(\lambda), \qquad {B}_l(\lambda)=R_l(\lambda)J_{l,+}(\lambda), 
\end{equation}
and for ${B}_D$ we have two formulas
\begin{equation}\label{bdrj}
    {B}_D(\lambda)=R_r(\lambda)J_{r,-}(\lambda)=R_l(\lambda)J_{l,-}(\lambda).
\end{equation}
The last equation defines a factorization problem 
\begin{equation}\label{factorprob2}
    G_{\mathcal{C}}(\lambda)=J_{l,-}(\lambda)\left(J_{r,-}(\lambda)\right)^{-1}=\left(R_l(\lambda)\right)^{-1}R_r(\lambda).
\end{equation}
Solutions for this factorization problem $R_{r}$ and $R_{l}$ can be multiplied only by a constant matrix from the left, which can be fixed using asymptotics near $\lambda=\infty$.

\subsection{Notion of degree of approximation}
To develop the perturbation theory systematically, we need a notion of degree of approximation. 
It is related directly to the number of terms in $G_0(\zeta)$ defined in \eref{G0def} which we keep, 
namely for $n$-th order of approximation we use the approximation
\begin{equation}
    G_0(\zeta)\approx I+\sum_{j=1}^n \frac{\rho_j}{\zeta^j}.
\end{equation}
The related function $G_r(\zeta)$ defined in \eref{Grdef} and its series \eref{epsilondef} motivate  the following definition  of degree of approximation on monomials
\begin{equation}\label{defdeg}
    \mathrm{deg}\left(x^{-\frac{k}{2}}\right)=k, \quad \mathrm{deg}\left(x^{-\frac{k}{2}}\tilde{Q}^{\pm s}\right)=k-s+1, \quad s>0.
\end{equation}
Then the $n$-th order of approximation can be defined as one for which we neglect the terms of degree greater than $n$. 
Note that two monomials in \eqref{defdeg} having the same degree may have or have not different asymptotic behaviors for large $x$, 
if we use $\tilde Q$ defined by \eqref{Qtilde}, depending on the value of $\nu$. 
However, we want to collect all such terms in order to have approximations which are independent of the value of parameter $\nu$.  

The degree of matrix-valued function is defined as the lowest degree of its matrix elements.
In particular,
\begin{equation}
    \mathrm{deg}\, \varepsilon_k=k, 
\qquad    
\mathrm{deg}\, (\varepsilon_j \varepsilon_k) =j+k.
\end{equation}

\subsection{$0$-order of approximation of gauge transformations}
It is hard to tackle the factorization problem \eref{factorprobdef} in general, thus we will find a solution perturbatively for large parameter $x$. 
We start with $0$-order approximation of $G_r(\lambda)$ which corresponds to the leading term of \eref{epsilondef} for large $x$:
\begin{equation}
    G_r(\lambda)\approx I.
\end{equation}
In this case, the jump for factorization problem \eref{Jrfactor} is rather simple 
\begin{equation}
    J_r=\beta_r(\lambda) G_r(\lambda)^{-1}\approx \beta_r(\lambda),
\end{equation}    
where we used \eref{Jrdef}.
Since this function is already belongs to $\mathcal{H}(D_r)$, we immediately read off $0$-order approximation for gauge transformations
\begin{equation}\label{br0def}
    B_D^{(0)}(\lambda)=I, \qquad B_r^{(0)}(\lambda)=\beta_r(\lambda).
\end{equation}
Due to $Z_2$-symmetry relations
\begin{equation}
    Y_D(\lambda)=\sigma_2 Y_D(-\lambda)\sigma_2, \qquad \Psi_l(\lambda)=\sigma_2 \Psi_r(-\lambda)\sigma_2,
\end{equation}
the factorization of the jump on the right circle leads to factorization of the jump on the left circle in the same order of approximation. Therefore we obtained a $0$-order approximation for the global solution of RH problem
having the expansion for large $\lambda$
\begin{equation}
    \Psi_D^{(0)}=Y_D(\lambda)=I+\frac{(\lambda_--\lambda_+)\nu\sigma_3}{\lambda}+O(\lambda^{-2}).
\end{equation}
It allows us to find the corresponding approximation for the Hamiltonian using \eref{Hamhatdef}
\begin{equation}
    \hat{H}^{(0)}=-\frac{i}{2}\Tr \hat{g}^{(0)}_1 \sigma_3 = i\nu.
\end{equation}
Integrating the Hamiltonian $\hat{H}^{(0)}$, we get leading asymptotic term for the tau function
\begin{equation}
    \tau^{(0)}=e^{i\nu s}.
\end{equation}

\section{Factorization problems on circles $\mathcal{C}_r$ and $\mathcal{C}_l$: $n$-th order of approximation }

\subsection{General construction}

As a first step, we want to solve the factorization problem \eref{Jrfactor} on $\mathcal{C}_r$ in $n$-th order of approximation with respect the degree defined above. We can rewrite it as
\begin{equation}
    J_{r,-}(\lambda) J_r(\lambda)= J_{r,+}(\lambda).
\end{equation}
Since $J_{r,+}(\lambda)\in\mathcal{H}(D_r)$, we have
\begin{equation}\label{holomorphcond}
    \left[ J_{r,-}(\lambda) J_r(\lambda)\right]_{-}=0,
\end{equation}
where we denote by $[...]_-$  the projection  onto the space of holomorphic functions outside $D_r$ which vanish at $\lambda=\infty$. 
This projection acts on each matrix elements.
The equation \eref{holomorphcond} rewritten in basis $(\lambda-\lambda_+)^k$, $k\in \mathbb{Z}$, becomes a system of linear equation for unknown 
Fourier components of  $J_{r,-}(\lambda)$  with respect to the basis. Let us describe it explicitly.

In $n$-th order of approximation, the jump matrix  $J_r(\lambda)$ on $\mathcal{C}_r$  is given by
\begin{equation}
    J_r(\lambda)=\beta_r(\lambda)G_r(\lambda)^{-1}, \qquad G_r(\lambda)^{-1}=I-\sum_{j=1}^n \frac{\eta_j}{w_+^j(\lambda)},
\end{equation}
where $ G_r(\lambda)^{-1}$ is obtained by inversion of 
\begin{equation}
    G_r(\lambda)=I+\sum_{j=1}^n \frac{\varepsilon_j}{w_+^j(\lambda)}
\end{equation}
neglecting terms of degree higher than $n$. 
Explicit expressions for the first coefficients of $ G_r(\lambda)^{-1}$  are
\begin{equation}
    \eta_1=\varepsilon_1, \qquad \eta_2=\varepsilon_2-\varepsilon_1^2, \qquad \ldots.  
\end{equation}
Using the definition of $w_+(\lambda)$ given by \eref{wpdef}, we see that the most singular term of expansion of $J_r(\lambda)$ at $\lambda=\lambda_+$ 
is $(\lambda-\lambda_+)^{-n}$. 
Therefore the Fourier series for $J_r(\lambda)$ on $\mathcal{C}_r$ has the form
\begin{equation}
    J_r(\lambda)=I-\sum_{k=-n}^{\infty} q_k (\lambda-\lambda_+)^k,
\end{equation}
where 
\begin{equation}
    q_k=\oint_{\mathcal{C}_r}\frac{d\lambda}{2\pi i}(\lambda-\lambda_+)^{-k-1}\left(I-J_r(\lambda)\right).
\end{equation}
Also we expect that a proper anzats for $J_{r,-}$ is
\begin{equation}\label{JrmE}
    J_{r,-}(\lambda)=I+\sum_{j=1}^{n}\frac{E_j}{(\lambda-\lambda_+)^j}.
\end{equation}
Now equation \eref{holomorphcond}, describing factorization of $J_r$, can be rewritten as a system of $2n$ matrix linear equations 
for unknown matrices $E_k$
\begin{equation}\label{Toeplitz1trivial}
    \sum_{k=j}^n E_k q_{k-j-n}=0, \qquad j=1, 2, \ldots, n,
\end{equation}
\begin{equation}\label{Toeplitz1}
    (E_n, \ldots, E_1)\begin{pmatrix}
		I-q_0 & -q_1 & \cdots & -q_{n-1}\\
		-q_{-1}& I-q_0 & \cdots & -q_{n-2}\\
		\vdots & \ddots & \ddots & \vdots\\
		-q_{-n+1} & \cdots & -q_{-1} & I-q_0
	\end{pmatrix}=(q_{-n},\ldots, q_{-1}).
\end{equation}
It is shown in \ref{AppCr} how to solve \eref{Toeplitz1} effectively and that $n$ matrix equations \eref{Toeplitz1trivial} are satisfied for the solution of equations \eref{Toeplitz1}.

Factorization of the jump $J_l$ on $\mathcal{C}_l$  can be done similarly. Since our construction of $J_l$ uses $Z_2$-symmetry we can use it to
find $J_{l,-}(\lambda)$  by 
\begin{equation}
    J_{l,-}(\lambda)=I +\sum_{j=1}^{n}\frac{\tilde E_j}{(\lambda-\lambda_-)^j}, \qquad  \tilde E_j=(-1)^j \sigma_2 E_j \sigma_2.
\end{equation}

\subsection{An example: the factorization in $n=1$ order of approximation}
\label{Example1}

In this subsection, we apply the described procedure to $n=1$ case.
Using \eref{Jrdef} we get the first order of approximation for $J_r$ 
\begin{equation}
    J_r(\lambda)=\beta_r(\lambda)\left( I-\frac{\varepsilon_1}{w_+(\lambda)}\right),
\end{equation}
where  we took into account $\varepsilon_1^2=0$ in this order of approximation to find the inverse of 
\begin{equation}
    G_r(\lambda)=I+\frac{\varepsilon_1}{w_+(\lambda)}.
\end{equation}
To find a solution $J_{r,-}(\lambda)$ of \eref{holomorphcond}, we assume that in the first order of approximation 
\begin{equation}
   J_{r,-}(\lambda)=I+\frac{E_1}{\lambda-\lambda_+}.
\end{equation}
The equations \eref{Toeplitz1trivial} and  \eref{Toeplitz1} are a system of two linear matrix equations for unknown matrix $E_1$
\begin{equation}\label{system1}
    E_1 q_{-1}=0, \qquad E_1(I-q_0)=q_{-1},
\end{equation}
where 
\begin{equation}
    q_{-1}=\varepsilon_1, \qquad q_0=(\alpha_{r,1}\cdot I + \beta_{r,1})\varepsilon_1
\end{equation}
with the coefficients $\alpha_{r,1}$ and $\beta_{r,1}$ defined by
\begin{equation}\label{Fourierbeta}
    \frac{\lambda-\lambda_+}{w_+(\lambda)}=1+\sum_{j=1}^{\infty}\alpha_{r,j}(\lambda-\lambda_+)^j, \quad
    \beta_r(\lambda)=I+\sum_{j=1}^{\infty}\beta_{r,j}(\lambda-\lambda_+)^j.
\end{equation}
Since the function $\beta_r(\lambda)$ defined in \eref{Grdef} is a diagonal matrix function we have $\varepsilon_1 q_0 = 0$ in the first-order approximation.
Therefore (all the equalities are up to terms of $\mathrm{deg}\,(.)>1$) 
\begin{equation}
    E_1=q_{-1}(I-q_0)^{-1}=\varepsilon_1 (I-q_0)^{-1}=\varepsilon_1.
\end{equation}
Note that the equation $ E_1 q_{-1}=0$ for $E_1$ is satisfied automatically. 

The factorization problem for $J_l$ can be solved similarly
\begin{equation}
    J_{l,-}(\lambda)=I+\frac{\tilde{E}_1}{\lambda-\lambda_-}, \qquad 
    \tilde{E}_1=\tilde{\varepsilon}_1=-\sigma_2 \varepsilon_1 \sigma_2.
\end{equation}
We finish the computation in subsection~\ref{Example1end} after describing the general procedure of the next step.

\section{Gluing local parametrices}

\subsection{General construction}

The second step is to solve the factorization problem \eref{factorprob2} rewritten as
\begin{equation}
    R_{l}(\lambda) G_{\mathcal{C}}(\lambda)=R_r(\lambda).
\end{equation}
Since the function $R_r(\lambda)\in\mathcal{H}(D_r)$, we have 
\begin{equation}\label{holomorphcond1}
    \left[R_{l}(\lambda) G_{\mathcal{C}}(\lambda)\right]_-=0,
\end{equation}
where as before 
we denote by $[...]_-$  the projection  onto the space of holomorphic functions outside $D_r$ which vanish at $\lambda=\infty$.
Using \eref{JrmE}, $\mathrm{deg}\,E_j=j$, and $\mathrm{deg}\,(E_jE_k)=j+k$, we have in $n$-order of approximation
\begin{equation}
	J_{r,-}(\lambda)^{-1}=I-\sum_{j=1}^{n}\frac{H_j}{(\lambda-\lambda_+)^j},
\end{equation}
where  $\mathrm{deg}\,H_j=j$. Therefore $G_\mathcal{C}(\lambda)$ can be presented as a series
\begin{equation}
    G_\mathcal{C}(\lambda)=I-\sum_{k=-n}^\infty f_k (\lambda-\lambda_+)^k
\end{equation}
with the Fourier components 
\begin{equation}\label{fourierF}
    f_k=\oint_{\mathcal{C}_r}\frac{d\lambda}{2\pi i}(\lambda-\lambda_+)^{-k-1}\left(I-G_{\mathcal{C}}(\lambda)\right).
\end{equation}
All these series structures lead to a natural anzats 
\begin{equation}\label{Rldef}
    R_{l}(\lambda)=I+\sum_{j=1}^n \frac{r_j}{(\lambda-\lambda_+)^j}.
\end{equation}

The equation \eref{holomorphcond1} rewritten in terms of Fourier components become a system of $2n$  linear matrix equations for uknown matrices $r_j$
\begin{equation}\label{Toeplitztrivial2}
    \sum_{k=j}^n r_k f_{k-j-n}=0, \qquad j=1, 2, \ldots, n,
\end{equation}
\begin{equation}\label{Toeplitz2}
    (r_n, \ldots, r_1)\begin{pmatrix}
		I-f_0 & -f_1 & \cdots & -f_{n-1}\\
		-f_{-1}& I-f_0 & \cdots & -f_{n-2}\\
		\vdots & \ddots & \ddots & \vdots\\
		-f_{-n+1} & \cdots & -f_{-1} & I-f_0
	\end{pmatrix}=(f_{-n},\ldots, f_{-1}).
\end{equation}
In \ref{AppGluing},  it  is shown that the solution of equations \eref{Toeplitz2} satisfies 
\eref{Toeplitztrivial2}. 
Using \eref{Rldef}, this solution defines $R_l(\lambda)$ which in turn gives $n$-order of approximation of $B_D(\lambda)$ and $\Psi_D(\lambda)$ due to the relations \eref{bdrj} and \eref{localparamD}.

Solving equations \eref{Toeplitz1} and \eref{Toeplitz2} we can find the Hamiltonian in the $n$-th order approximation given by the formula \eref{Hamhatdef}, namely
\begin{equation}
    \hat{H}^{(n)}=i\nu - \frac{i}{2}\Tr\left(\tilde{E}_1+r_1\right)\sigma_3,
\end{equation}
where $\tilde{E}_1$ is related to $E_1$ by the formula
\begin{equation}
    \tilde{E}_1=-\sigma_2 E_1 \sigma_2.
\end{equation}

\subsection{The final step of the factorization in $n=1$ order of approximation }
\label{Example1end}

In this subsection, we finish computations for $n=1$ case started in subsection~\ref{Example1}.
Assuming that in the first order of approximation the function $R_l(\lambda)$ has the form
\begin{equation}
    R_l(\lambda)=I+\frac{r_1}{\lambda-\lambda_+},
\end{equation}
the equation \eref{holomorphcond1} leads to equations 
\begin{equation}\label{systemR}
    r_1 f_{-1}=0, \qquad r_1(I-f_0)=f_{-1}.
\end{equation}
Using \eref{fourierF} we get
\begin{equation}
    f_{-1}=(I+\tilde{\varepsilon}_1)\varepsilon_1, \qquad f_0=-\tilde{\varepsilon}_1(I+\varepsilon_1).
\end{equation}
The solution of the system \eref{systemR} is
\begin{equation}
    r_1=f_{-1}(I-f_0)^{-1}.
\end{equation}
The first equation of \eref{systemR} is satisfied automatically due to 
\begin{equation}\label{M2deg2}
    (I-f_0)^{-1}f_{-1}=(I+(I+\tilde{\varepsilon}_1)^{-1}\tilde{\varepsilon}_1 \varepsilon_1)\varepsilon_1=\varepsilon_1, \qquad f_{-1}\varepsilon_1=0.
\end{equation}
Here we used $\varepsilon_1^2=0$ in this approximation. Using also $\tilde{\varepsilon}_1^2=0$, we can present $r_1$ in more convenient form
\begin{equation}\label{r1approx1}
    r_1=\tilde{\varepsilon}_1\varepsilon_1 \Delta^{-1}-\varepsilon_1 \tilde{\varepsilon}_1 \tilde{\Delta}^{-1}+\varepsilon_1 \Delta^{-1}-\tilde{\varepsilon}_1\varepsilon_1 \Delta^{-1}\tilde{\varepsilon}_1,
\end{equation}
where we introduced matrices
\begin{equation}
    \Delta=I+\tilde{\varepsilon}_1\varepsilon_1, \qquad \tilde{\Delta}^{-1}=I+\varepsilon_1 \tilde{\varepsilon}_1.
\end{equation}
Note that the first two matrices in formula \eref{r1approx1} are diagonal and the last two are off-diagonal. Let  us introduce the following notations
\begin{equation}\label{Vpm1}
    V_1=\frac{\tilde{Q}^2 \nu^2}{c_+^2 x}, \qquad V_{-1}=\frac{1}{\tilde{Q}^2 c_+^2 x},
\end{equation}
then, using $\varepsilon_1$ given by \eref{epsilondef} and $\tilde{\varepsilon}_1=-\sigma_2 \varepsilon_1 \sigma_2$,  we  have 
\begin{equation}
    \tilde{\varepsilon}_1\varepsilon_1=
    \begin{pmatrix}
        V_{-1} & 0\\
        0 & V_1
    \end{pmatrix}, \qquad 
    \varepsilon_1 \tilde{\varepsilon}_1=
    \begin{pmatrix}
        V_{1} & 0\\
        0 & V_{-1}
    \end{pmatrix},
\end{equation}
and the diagonal part of $r_1$ is
\begin{equation}
    r_1^{(\mathrm{diag})}=-\left(\frac{V_1}{1+V_1}-\frac{V_{-1}}{1+V_{-1}}\right)\sigma_3.
\end{equation}
Using $\Psi_D$ and $B_D$ given by \eref{localparamD} and \eref{bdrj}, we obtain the expansion of $\Psi_D$ at $\lambda=\infty$
\begin{equation}
    \Psi_D(\lambda)=I+\frac{1}{\lambda}\left(-\nu\sigma_3 + \tilde{\varepsilon}_1+r_1\right)+O(\lambda^{-2}).
\end{equation}
Thus, in the first-order approximation, using \eref{Hamhatdef}, we get
\begin{equation}
    \hat{H}^{(1)}=i\nu + i\left(\frac{V_1}{1+V_1}-\frac{V_{-1}}{1+V_{-1}}\right).
\end{equation}
The denominators $1+V_{\pm 1}$ were neglected in \cite{fokas2006painleve}, but they are important 
for the asymptotic analysis of tau functions. We were inspired by \cite{Its_2018} where similar denominators were 
derived for Painlev\'e~VI. Note, all the terms in expansion of these denominators into geometric series have the same degree 
with respect to \eqref{defdeg}.

This Hamiltonian can be integrated using the fact that up to terms of second degree we have the relation
\begin{equation}
    \partial_s \log(1+V_1 + V_{-1})=\frac{i(V_1-V_{-1})}{1+V_1+V_{-1}}=i\left(\frac{V_1}{1+V_1}-\frac{V_{-1}}{1+V_{-1}}\right).
\end{equation}
Therefore the first-order approximated tau function is
\begin{equation}
    \tau^{(1)}=e^{i\nu s}(1+V_1 + V_{-1}).
\end{equation}

Similarly, using this algorithm for $n=2$ we obtain 
\begin{equation}
    \hat{H}^{(2)}=i\nu + i\left(\frac{V_1}{1+V_1}-\frac{V_{-1}}{1+V_{-1}}\right)-\frac{\nu^2}{s}.
\end{equation}
Neglecting the terms of degree higher than $2$ we can integrate this expression to get
\begin{equation}
    \tau^{(2)}=e^{i\nu s}s^{-\nu^2}\left(1+V_{1}+V_{-1}\right).
\end{equation}
This expression coincides with the second order of approximation of \eref{tausum} (up to inessential multiplier
$C(\nu)$) due to the fact that $V_{\pm 1}$ defined by \eref{Vpm1} can be presented also as 
\begin{equation}
    V_k=\frac{C(\nu+k)}{C(\nu)} \frac{e^{i(\nu+k) s}s^{-(\nu+k)^2}}{e^{i\nu s}s^{-\nu^2}} e^{ik\rho}.
\end{equation}

\section{Discussion}

 In this paper we elaborated a systematic way to calculate higher orders of asymptotic expansion series 
 for the Hamiltonian of Painlev\'e~II equation in the long-time limit $t\to -\infty$.
We believe that our approach can be applied  to asymptotic analysis of different RH problems and, in particular,
RH problems associated with the other Painlev\'e equations. In a forthcoming paper, the described method with minor changes will be applied 
for higher order asymptotic analysis of Painlev\'e~I equation.

Also this method gives asymptotic series for the isomonodromic tau function by integration of the Hamiltonian. 
However this approach looks a bit indirect and it is interesting to develop a method of obtaining tau function directly.
Such a direct way is known for tau function of Painlev\'e~VI equation where this tau function is presented 
as a Fredholm determinant related to the Widom constant \cite{Gavrylenko_2018,Cafasso_2018}.
We plan to construct a similar Fredholm determinant presentation of tau function of Painlev\'e~II equation
in the form of a generalization of  Widom constant on two circles $\mathcal{C}_r$ and $\mathcal{C}_l$ using 
the technique of the present paper.
Also we would like to mention recent paper \cite{Desiraju_2021}, where the author 
found a Fredholm determinant presentation for Painlev\'e~II tau function, but it seems that this construction is not related 
apparently to Widom constant. In another recent paper \cite{hao2025newconstructionc1virasoro}, a general CFT approach of 
Fredholm determinant construction of tau functions was suggested. 

Recently a new approach to construct solutions of Painlev\'e equations was developed on the base of topological recursion
on elliptic curves \cite{Iwaki2020,Marchal2022,IILZ2025}. This approach is closely related to holomorphic anomaly equation approach 
\cite{Fucito2023,bonelli2024surfaceobservablesgaugetheories,bonelli2025refinedpainlevegaugetheorycorrespondence,IM2024} allowing to compute effectively Painlev\'e tau functions.

One more direction of research is motivated by the relation of  Painlev\'e equations  with conformal field theory 
\cite{Gamayun2012,Gamayun2013,Iorgov2014}. In this direction,  the most challenging  problem is to 
build irregular counterparts of conformal blocks together with their relation to Painlev\'e equations
\cite{Gaiotto2012,Lisovyy2018,Nishinaka2019,Poghosyan2023,Hamachika2024,poghossian2025note32,IILZ2025}.

\ack 
We are grateful to Oleksandr Gamayun, Pavlo Gavrylenko,  Harini Desiraju,  Oleg Lisovyy, Mykola Semenyakin, and Piotr Su\l kowski  for useful discussions.
The authors acknowledge support by the National
Research Foundation of Ukraine grant 2020.02/0296, the Simons Foundation, and NAS of Ukraine (project No. 0122U000888).

\appendix

\section{Details of factorization on $\mathcal{C}_r$}
\label{AppCr}

In this appendix we present an effective way to invert the block Toeplitz matrix
\begin{equation}
\hat q = 
\begin{pmatrix}
		I-q_0 & -q_1 & \cdots & -q_{n-1}\\
		-q_{-1}& I-q_0 & \cdots & -q_{n-2}\\
		\vdots & \ddots & \ddots & \vdots\\
		-q_{-n+1} & \cdots & -q_{-1} & I-q_0
	\end{pmatrix}
\end{equation}
defining the coefficients of the linear system \eref{Toeplitz1}. Also we show that 
the equations \eref{Toeplitz1trivial} are satisfied automatically on the base of degree reasoning.

To invert $\hat q$, it is convenient to present it as 
\begin{equation}
    \hat{q}=\delta+\hat{q}^{(1)},
\end{equation}
where $\delta$ is its part of degree $0$ and by $\hat{q}^{(1)}$ the remaining part. 
The matrix $\delta$ can be expressed in terms of Fourier coefficients \eref{Fourierbeta} 
of the matrix $\beta_r(\lambda)$ defined in \eref{Grdef}  as
\begin{equation}
    \delta=\begin{pmatrix}
		I & \beta_{r,1} & \cdots & \beta_{r, n-1}\\
		0 & I & \cdots & \beta_{r, n-2}\\
		\vdots & \ddots & \ddots & \vdots\\
		0 & \cdots & 0 & I
	\end{pmatrix}.
\end{equation}
This matrix can be easily inverted even for generic $n$ giving an effective formula for $\hat{q}^{-1}$ 
\begin{equation}
    \hat{q}^{-1}=\delta^{-1}\left(I+\hat{q}^{(1)}\delta^{-1}\right)^{-1}=\delta^{-1}\sum_{j=0}^n \left(-\hat{q}^{(1)}\delta^{-1}\right)^j
\end{equation}
valid in the $n$-th order of approximation.
Using this formula and equation \eref{Toeplitz1} we obtain
\begin{equation}
    \mathrm{deg}\, E_k=\mathrm{deg}\,q_{-k}=k, \qquad k>0.
\end{equation}
Taking into account
\begin{equation}
    \mathrm{deg}\, E_j q_{-k}=j+k,
\end{equation}
we obtain that the equations \eref{Toeplitz1trivial} are satisfied automatically for the solution of equations \eref{Toeplitz1}. 

\section{Details of factorization needed for gluing local parametrices}
\label{AppGluing}

In this appendix we show how to solve the system \eref{Toeplitztrivial2} and \eref{Toeplitz2}.
It turns out the solution of equations \eref{Toeplitz2} automatically  satisfies 
\eref{Toeplitztrivial2}.

\subsection{Some properties of Fourier coefficients $f_k$}

First, we give some information on the Fourier coefficients  $f_k$ entering the equations \eref{Toeplitztrivial2} and \eref{Toeplitz2}.   
 Using explicit form of matrices $f_k$ we can obtain an estimation on their degrees. For the negative coefficients we have
\begin{equation}
    \mathrm{deg}\, f_{-k}= k, \qquad k>0,
\end{equation}
and for the  non-negative coefficients we have
\begin{equation}
    \mathrm{deg}\, f_{k}= 1, \qquad k\geq 0.
\end{equation}
There is an inequality for degrees having the form
\begin{equation}\label{ineq1}
    \mathrm{deg}\,AB \geq \mathrm{deg}\, A+\mathrm{deg}\, B - 1,
\end{equation}
which can be proven using monomials, namely, the worst case is the case of two monomials having the same signs of exponents of $\tilde{Q}$, for example
\begin{equation}
    \mathrm{deg}\, x^{-\frac{k_1}{2}}\tilde{Q}^{s_1}\cdot x^{-\frac{k_2}{2}}\tilde{Q}^{s_2}= k_1+k_2-(s_1+s_2)+1= \mathrm{deg}\, x^{-\frac{k_1}{2}}\tilde{Q}^{s_1} + \mathrm{deg}\, x^{-\frac{k_2}{2}}\tilde{Q}^{s_2}-1.
\end{equation}
We can find degree of matrix $r_n$ from the first equation of the system \eref{Toeplitz2}
\begin{equation}
    r_n(1-f_0)=\left(r_{n-1}f_{-1}+\ldots+ r_1f_{-n+1}\right)+f_{-n} \qquad \Rightarrow \qquad \mathrm{deg}\,r_n=n.
\end{equation}
The implication is true because $\mathrm{deg}\,f_{-n}=n$ and the expression in brackets has also degree $n$ by induction. 
We also need to fulfil \eref{Toeplitztrivial2}. Using the inequality \eref{ineq1} we get 
\begin{equation}
    \mathrm{deg}\left(\sum_{k=j}^n r_k f_{k-j-n}\right)\geq n+j-1.
\end{equation}
Thus, for $j>1$, the equations  \eref{Toeplitztrivial2} are satisfied automatically. The only equation which we need to 
prove is
\begin{equation}\label{condMnp1}
    M_{n+1}:=r_n f_{-1}+r_{n-1}f_{-2}+\cdots +r_1 f_{-n}=0
\end{equation}
in $n$-order of approximation.

We will prove that 
\begin{equation}
	\mathrm{deg}\, M_{n+1}=n+1,
\end{equation}
where the matrices $r_k$ solve the system of equation \eref{Toeplitz2}
\begin{equation}\label{systemMk}
	\begin{cases}
		r_n(1-f_0)-M_n=f_{-n},\\
		-r_n f_1 + r_{n-1}(1-f_0)-M_{n-1}=f_{-n+1},\\
		\vdots\\
		-r_n f_{n-1}-\cdots - r_2 f_1 + r_1 (1-f_0)=f_{-1}.
	\end{cases}
\end{equation}
We already have seen that $\mathrm{deg}\, M_2=2$ for the $n=1$ case, see \eref{M2deg2}, therefore we will assume that $\mathrm{deg}\, M_k=k$ for $k<n+1$.
\subsection{Main tricks}
We want to present some properties of matrices $f_k$ following from their definition \eref{fourierF}. 
Let we define the matrices $H_j$ as
\begin{equation}
	J_{r,-}^{-1}=\left(1+\sum_{j=1}\frac{E_j}{(\lambda-\lambda_+)^j}\right)^{-1}=1-\sum_{j=1}^{\infty}\frac{H_j}{(\lambda-\lambda_+)^j}.
\end{equation}
These matrices $H_j$ have the same dependence on $\tilde{Q}$ as $E_j$ (and $\varepsilon_j$), namely
\begin{equation}
	E_j=\tilde{Q}^{\sigma_3/2}e_j \tilde{Q}^{-\sigma_3/2}, \qquad H_j=\tilde{Q}^{\sigma_3/2}h_j \tilde{Q}^{-\sigma_3/2},
\end{equation}
where matrices $e_j, h_j$ are independent of $\tilde{Q}$. Therefore, we can easily find the degree of a product of $H_j$'s
\begin{equation}
	\mathrm{deg}\left(H_{j_1}\ldots H_{j_k}\right)=j_1+\cdots+j_k.
\end{equation}
It is convenient to introduce a vector space $\boldsymbol{H}_l$ generated by the following monomials
\begin{equation}
	\boldsymbol{H}_l=\mathrm{Span}\,(H_{j_1}\ldots H_{j_k}, \ j_1+\cdots+j_k=l).
\end{equation}
Let we discuss the leading form of $f_{-k}$
\begin{equation}\label{fklead}
	f_{-k}=f_{-k}^{(k)} + \delta f_{-k}, \qquad f_{-k}^{(k)}=(1+\tilde{\varepsilon}_1)H_k,
\end{equation}
where $\delta f_{-k}$ is some matrix having degree $k+1$. Also, we will need the leading behavior of $f_0$
\begin{equation}
	f_0=f_0^{(1)}+\delta f_0, \qquad f_0^{(1)}=-\tilde{\varepsilon}_1(1+\varepsilon_1),
\end{equation}
where $\delta f_0$ is a matrix of degree 2. The precise form of $f_{-k}$ and $f_0$ allow us to prove a useful relation
\begin{equation}\label{relationHH}
	(1-f_0)^{-1}f_{-k}\boldsymbol{H}_{n-k}=H_k \boldsymbol{H}_{n-k},
\end{equation}
where we droped terms of degree higher than $n$. Using the inequality for degrees \eref{ineq1} we can drop $\delta f_0$ and $\delta f_{-k}$ to get
\begin{multline}
	(1-f_0)^{-1}f_{-k}\boldsymbol{H}_{n-k}=(1+\tilde{\varepsilon}_1+\tilde{\varepsilon}_1\varepsilon_1)^{-1}(1+\tilde{\varepsilon}_1)H_k \boldsymbol{H}_{n-k}\\
        =(1+(1+\tilde{\varepsilon}_1)^{-1}\tilde{\varepsilon}_1\varepsilon_1)^{-1}H_k \boldsymbol{H}_{n-k}.
\end{multline}
Using the fact that
\begin{equation}
	\mathrm{deg}\, \varepsilon_1 H_k \boldsymbol{H}_{n-k}=n+1,
\end{equation}
we obtain the relation \eref{relationHH}.
\subsection{Proof of \eref{condMnp1}}
Let we show that 
\begin{equation}
	\mathrm{deg}\,r_1 f_{-n}=n+1.
\end{equation}
We can find $r_1$ from the system \eref{systemMk} in the form
\begin{equation}\label{exprr1}
	r_1=(f_{-1}+r_2 f_1+\cdots+r_n f_{n-1})(1-f_0)^{-1}.
\end{equation}
Due to the fact that $f_{-n}$ has degree $n$ we can drop terms having degree higher than 1 
\begin{equation}
	r_1 f_{-n}=f_{-1}(1-f_0)^{-1}f_{-n}=f_{-1}H_n=0,
\end{equation}
where we used the identity \eref{relationHH} for $k=n$ and the last equality follows from the leading behavior of $f_{-1}$ given by \eref{fklead}. Let we show that
\begin{equation}
	\mathrm{deg}\,r_2 f_{-n+1}=n+1.
\end{equation}
We can find $r_2$ from the system \eref{systemMk} in the form
\begin{equation}
	r_2=(f_{-2}+M_2+r_3 f_1+\cdots+r_n f_{n-2})(1-f_0)^{-1}.
\end{equation}
Due to the fact that $f_{-n+1}$ has degree $n-1$ we can drop terms having degree higher than 2
\begin{equation}
	r_2 f_{-n+1}=(f_{-2}+M_2)(1-f_0)^{-1}f_{-n+1}.
\end{equation} 
In the last expression, only leading terms of $f_k$ contribute
\begin{equation}
 (f_{-2}+M_2)(1-f_0^{(1)})^{-1}f_{-n+1}^{(n-1)}=(f_{-2}+M_2)H_{n-1}=M_2 H_{n-1}.
\end{equation}
On this step, we need to use expression for $r_1$ given by \eref{exprr1}
\begin{equation}
	M_2 H_{n-1}=r_1 f_{-1}H_{n-1}=f_{-1}(1-f_0)^{-1}f_{-1}H_{n-1}.
\end{equation}
Using the identity \eref{relationHH} for $k=1$, we get
\begin{equation}
	M_2 H_{n-1}=f_{-1}H_1 H_{n-1}=0,
\end{equation}
where we used the leading behavior of $f_{-1}$ given by \eref{fklead}. We can repeat this logic for $r_k$. We want to show that 
\begin{equation}
	\mathrm{deg}\,r_k f_{-n+k-1}=n+1.
\end{equation}
We can find $r_k$ from the system \eref{systemMk} in the form
\begin{equation}\label{exprrk}
	r_k=(f_{-k}+M_k+r_{k+1} f_1+\cdots+r_n f_{n-k})(1-f_0)^{-1}.
\end{equation}
Due to the fact that $f_{-n+k-1}$ has degree $n-k+1$ we can drop terms having degree higher than $k$, therefore we have
\begin{equation}\label{statement1}
	r_k f_{-n+k-1}=(f_{-k}+M_k)(1-f_0)^{-1}f_{-n+k-1}.
\end{equation}
In the last expression, only leading terms of $f_j$ contribute
\begin{equation}
	(f_{-k}+M_k)(1-f_0^{(1)})^{-1}f_{-n+k-1}^{(n-k+1)}=(f_{-k}+M_k)H_{n-k+1}=M_k H_{n-k+1}.
\end{equation}
In this way, we reduced the problem to a statement that if we have
\begin{equation}\label{statement2}
	\mathrm{deg}\, M_k \boldsymbol{H}_{n-k+1}=n+1, \qquad k<n+1,
\end{equation}
then we can prove the statement \eref{exprrk}.
The proof uses induction. Let we consider the $k=2$ case
\begin{equation}
	M_2 \boldsymbol{H}_{n-1}=r_1 f_{-1}\boldsymbol{H}_{n-1}.
\end{equation}
Using expression for $r_1$ given by \eref{exprr1} we have
\begin{equation}
	r_1 f_{-1}\boldsymbol{H}_{n-1}=f_{-1}(1-f_0)^{-1}f_{-1}\boldsymbol{H}_{n-1}=f_{-1}H_1 \boldsymbol{H}_{n-1}=0.
\end{equation}
Thus, we proved the base of induction. By similar arguments, we can prove that
\begin{equation}
	\mathrm{deg}\, r_1 f_{-k+1}\boldsymbol{H}_{n-k+1}=n+1.
\end{equation}
Indeed, we have
\begin{equation}
	r_1 f_{-k+1}\boldsymbol{H}_{n-k+1}=f_{-1}(1-f_0)^{-1}f_{-k+1}\boldsymbol{H}_{n-k+1}=f_{-1}H_{k-1}\boldsymbol{H}_{n-k+1}=0.
\end{equation}
For the other terms of $M_k$ we can formulate a similar statement
\begin{equation}
	\mathrm{deg}\, r_m f_{-k+m}\boldsymbol{H}_{n-k+m}=n+1, \qquad m<k.
\end{equation}
Using expression \eref{exprrk} for $k=m$, we get
\begin{equation}
	r_m f_{-k+m}\boldsymbol{H}_{n-k+1}=(f_{-m}+M_m)(1-f_0)^{-1}f_{-k+m}\boldsymbol{H}_{n-k+1}.
\end{equation}
Then, using the identity \eref{relationHH}, we have
\begin{equation}
	r_m f_{-k+m}\boldsymbol{H}_{n-k+1}=(f_{-m}+M_m)H_{k-m}\boldsymbol{H}_{n-k+1}=0,
\end{equation}
where the first term is zero due to the structure of leading behavior of $f_{-m}$ and the second term is zero by induction. This completes the proof of \eqref{statement2}, \eqref{statement1}, and finally \eref{condMnp1}.

\section*{References}

\bibliography{painleveII}

\end{document}